\documentclass{article}
\usepackage{spconf,amsmath,graphicx}
\usepackage{booktabs}
\usepackage{hyperref}
\usepackage{CJK}

\title{Findings of the 2024 Mandarin Stuttering Event Detection and Automatic Speech Recognition Challenge}
\name{\parbox{\linewidth}{\centering Hongfei Xue$^{1}$, Rong Gong$^{2}$, Mingchen Shao$^{1}$, Xin Xu$^{3}$, Lezhi Wang$^{2}$, Lei Xie$^{1 *}$\thanks{* Corresponding author.}, Hui Bu$^{3}$, Jiaming Zhou$^{4}$, Yong Qin$^{4}$, Jun Du$^{5}$, Ming Li$^{6}$, Binbin Zhang$^{7}$, Bin Jia$^{2}$}}
\address{
  $^1$ASLP@NPU, Northwestern Polytechnical University
  $^2$StammerTalk 
  $^3$AIShell Inc \\
  $^4$Nankai University
  $^5$University of Science and Technology of China \\
  $^6$Wuhan University
  $^7$WeNet Open Source Community}
  
\begin{document}
%\ninept
%
\maketitle
\begin{abstract}
The StutteringSpeech Challenge focuses on advancing speech technologies for people who stutter, specifically targeting Stuttering Event Detection (SED) and Automatic Speech Recognition (ASR) in Mandarin. The challenge comprises three tracks: (1) SED, which aims to develop systems for detection of stuttering events; (2) ASR, which focuses on creating robust systems for recognizing stuttered speech; and (3) Research track for innovative approaches utilizing the provided dataset. We utilizes an open-source Mandarin stuttering dataset AS-70, which has been split into new training and test sets for the challenge. This paper presents the dataset, details the challenge tracks, and analyzes the performance of the top systems, highlighting improvements in detection accuracy and reductions in recognition error rates. Our findings underscore the potential of specialized models and augmentation strategies in developing stuttered speech technologies.
\end{abstract}
\section{Introduction}
\label{sec:intro}
Stuttering is a speech impediment that affects approximately 1\% of the global population~\cite{yairi_1996}, characterized by disruptions such as repetitions, prolongations, and blocks. These disruptions significantly impact social interactions and mental well-being~\cite{stutteringoverview}, often leading to stress, shame, and low self-esteem for people who stutter (PWS). Consequently, PWS frequently experience communication avoidance and social isolation. Early intervention is crucial for effective treatment, particularly in children. However, regions like Mainland China face a shortage of certified speech therapists, limiting access to necessary support. With the rise of voice-user interfaces in smart home devices and chatbots like ChatGPT~\cite{openai2022chatgpt}, the need for inclusive speech technologies has become more pressing. However, current Automatic Speech Recognition (ASR) systems struggle with stuttering speech~\cite{Lea_chi_2023,Shonibare2022EnhancingAF, gong202470}, likely due to a lack of data. These problems highlight the need for stuttering event detection (SED) and inclusive speech technologies for PWS.

To address these issues, we organized the Mandarin Stuttering Event Detection and Automatic Speech Recognition (StutteringSpeech) Challenge, the first of its kind focused on Mandarin stuttering. The challenge aims to mobilize researchers to develop robust systems for detecting and recognizing stuttered speech, promoting the inclusion of PWS in the advancement of speech technologies.

The StutteringSpeech Challenge consists of three tracks:
Track I SED: This track focuses on developing systems capable of accurately different types of stuttering events in speech. Early detection is essential for timely intervention and treatment, making this track vital for improving the lives of PWS. 
Track II ASR: This track aims to create specialized ASR systems that can effectively recognize and transcribe stuttered speech. Current ASR systems struggle with stuttered speech, and this track encourages the development of more inclusive technologies. 
Track III Research: This track welcomes open submissions of papers on related topics.

We utilize the AS-70 dataset~\cite{gong202470}, the largest and most comprehensive open-source Chinese Mandarin stuttering dataset, as the challenge data.
The dataset includes conversational and voice command reading speech, providing a robust foundation for developing and testing SED and ASR systems. We have repartitioned the dataset to prevent overlap in command texts and avoid overly optimistic results in the command test sets.

This paper presents the findings of the StutteringSpeech Challenge, providing detailed insights into the dataset, track setup, and the performance of the submitted systems. Key findings from the challenge include: (1) The importance of tailored methods for different stuttered speech events, such as the Zipformer model~\cite{24zipformer} in handling word or phrase repetitions. (2) The effectiveness of specific data augmentation techniques for different types of stuttered speech events.

% By facilitating the development of inclusive speech technologies, this challenge seeks to improve the quality of life for PWS.

\section{Related Works}
The study of stuttering speech and its impact on speech technologies has gained increasing attention recently~\cite{Lea_chi_2023, Shonibare2022EnhancingAF, Bob_euphonia_2021, analysis-and-tuning, stutter-tts}. 
Most available datasets, such as FluencyBank~\cite{fluencybank} and UCLASS~\cite{uclass}, primarily focus on Western languages and are often small in size. The Sep-28k~\cite{sep-28k} dataset comprises 28,000 three-second podcast audio clips labeled for stuttering events but lacks text transcriptions. Another notable dataset is LibriStutter~\cite{libristutter} includes text transcriptions but is artificially generated from recordings of fluent speech.
Despite these efforts, there remains a significant resource gap for Mandarin-speaking PWS.
AS-70~\cite{gong202470} is a comprehensive dataset encompassing conversational and voice command reading speech, complete with manual transcriptions. For this challenge, we utilize the AS-70 dataset to provide a robust foundation for developing SED and ASR systems.

% However, the development of robust SED and ASR systems has been constrained by the lack of large-scale, openly accessible datasets, particularly for non-Western languages~\cite{fluencybank, uclass}. 
% Most available datasets, such as FluencyBank~\cite{fluencybank} and UCLASS~\cite{uclass}, primarily focus on Western languages and are often small in size, limiting their utility for developing comprehensive models. To address these limitations, several significant datasets have been introduced. 

Various neural network architectures have been explored for SED task. ConvLSTM~\cite{sep-28k} adds convolutional layers per feature type and learns how the features should be weighted. StutterNet~\cite{stutternet} uses a time-delay neural network~\cite{89tdnn} to capturing contextual aspects of the disfluent utterances. FluentNet~\cite{libristutter} consists of a Squeeze-and-Excitation Residual convolutional neural network followed by bidirectional long short-term memory layers~\cite{16blstm}, enhancing the learning of temporal relationships. Additionally, machine learning techniques like multi-task learning~\cite{sep-28k,bayerl23_interspeech} are employed to enhance the accuracy of stuttering detection. We use the conformer encoder as the baseline for the SED track~\cite{conformer}. The Conformer model, with its hybrid architecture combining convolutional operations and self-attention mechanisms, effectively captures both local and global dependencies in speech signals, making it particularly suitable for detecting nuanced stuttering events.

End-to-end ASR models such as Connectionist Temporal Classification (CTC)~\cite{20ctc}, recurrent neural network transducer (RNN-T)~\cite{12rnnt}, and attention-based encoder-decoder (AED)~\cite{15aed} have gained increasing attention over the last few years. Hybrid CTC/attention models~\cite{kim2017joint, 17hybrid, wenet} adopt both CTC and attention decoder loss during training, which results in faster convergence and improves the AED model's robustness. Additional, most ASR research on stuttered speech~\cite{Lea_chi_2023, Shonibare2022EnhancingAF, stutter-tts, Alharbi2017AutomaticRO} has primarily focused on predicting semantic content rather than exact word-for-word transcriptions. Therefore, in this study, we use a hybrid CTC/attention model u2++~\cite{21u2++} as the baseline for the ASR track. 

% U2++~\cite{21u2++} uses the forward and the backward information of the labeling sequences simultaneously during training to learn richer information.

% While previous studies have made significant strides, there remains a notable gap in resources and research for non-Western languages, particularly Mandarin. Our challenge aims to address this by providing a robust dataset and baseline systems, encouraging further innovation in the field of stuttering speech technologies.

\section{Dataset and Tracks}
\subsection{Dataset}
We use the AS-70 Mandarin dataset~\cite{gong202470} for this challenge. This dataset was gathered during 70 online voice chat sessions and comprised of conversation and voice command reading. Each recording session begins with an approximately half-hour interview conducted by two native Mandarin speakers who are both PWS. Following the interview, the interviewee is asked to read a prepared list of commands. The effective duration of the dataset is around 50 hours, encompassing recordings from 72 distinct PWS -- 2 interviewers and 70 interviewees.
% Gong et al.'s work

While AS-70~\cite{gong202470} has already partitioned the dataset into training, development, and test sets, we repartition the data according to specific distribution and stuttering severity of speakers detailed in Table~\ref{tab:speaker}. This repartitioning is essential because the original partitioning led to overlapping command texts between the training and test sets, although from different speakers, which could result in overly optimistic testing results for the command part.

\begin{table}[ht!]
\centering
\caption{Speaker numbers per stuttering severity and partition.}
\label{tab:speaker}
\begin{tabular}{@{}lllll@{}}
\toprule
      & Mild & Moderate & Severe & Sum \\ \midrule
Train & 25   & 12       & 6      & 43  \\
Dev   & 4    & 2        & 1      & 7   \\
Test  & 14   & 3        & 3      & 20  \\
Sum   & 43   & 17       & 10     & 70  \\ \bottomrule
\end{tabular}
\end{table}

For the SED task, five types of stuttering event~\cite{gong202470} are specified by the annotation guidelines, including:
\begin{CJK*}{UTF8}{gbsn}
\begin{itemize}
\item \textbf{/p}: \textbf{prolongation}. Elongated phoneme.
\item \textbf{/b}: \textbf{block}. Gasps for air or stuttered pauses.
\item \textbf{/r}: \textbf{sound repetition}. Repeated phoneme that do not constitute an entire character.
\item \textbf{[]}: \textbf{Word/phrase repetition}. Designated for marking entire repeated character or phrase.
\item \textbf{/i}: \textbf{interjections}. Filler characters due to stuttering e.g., `嗯', `啊', or `呃'. Notably, naturally occurring interjections that don't disrupt the speech flow are excluded.
\end{itemize}
\end{CJK*}
The number of utterances of each stuttering event type and the total utterance number are shown in Table~\ref{tab:sed_data}.

\begin{table}[ht!]
\centering
\caption{SED data on the number of utterances for the five stuttering types. Each utterance may contain multiple stuttering types or none at all.}
\label{tab:sed_data}
\begin{tabular}{@{}lllllll@{}}
\toprule
     & \textbf{/p} & \textbf{/b} & \textbf{/r} & \textbf{[]} & \textbf{/i}  & Sum\\ \midrule
Train    & 3,874   & 1,611       & 2,194     & 6,780 & 3,252  & 26,659 \\
Dev    & 450   & 203       & 286     & 1,435 & 591  & 4,294 \\
Test    & 1,208   & 486       & 770     & 2,527 & 1,424  & 11,000 \\
Sum    & 5,532   & 2,300       & 3,250     & 10,742 & 5,267  & 41,953 \\ \bottomrule
\end{tabular}
\end{table}

For the ASR task, the number of utterances contained in the training, development and test sets are shown in Table~\ref{tab:asr_data}. 
% We allow the use of AISHELL-1~\cite{aishell_2017} and acoustic data augmentation data, e.g., Musan\cite{15musan}, and RIR.

\begin{table}[ht!]
\centering
\caption{ASR data on the number of utterances in each partition.}
\label{tab:asr_data}
\begin{tabular}{@{}llll@{}}
\toprule
      Train & Dev & Test & Sum \\ \midrule
 22,866   & 3,903       & 8,559     & 35,328  \\ \bottomrule
\end{tabular}
\vspace{-10pt}
\end{table}

For a detailed description of this dataset, including the annotation process and statistical analysis, please refer to the papers by Gong et al. and Li et al.~\cite{gong202470, li2024towards}.

\subsection{Tracks}
The challenge comprises of three sub-tracks. For Track I and Track II, participants are restricted from using external audio data, timestamps, pre-trained models, and other information, except for AISHELL-1~\cite{aishell_2017} or acoustic data augmentation resources, such as Musan\cite{15musan} and RIR. Track III permits the use of any resources to improve results. Participants are encouraged to prioritize technological innovation, particularly through the exploration of novel model architectures, rather than relying solely on improvementd data usage.

\subsubsection{Track I-Stuttering Event Detection} 
This is a multi-label classification task. Participants are tasked with developing models to identify stuttering events in short speech audio snippets. The five types of stuttering events that may appear in the audio snippets are sound prolongation, sound repetition, character repetition, block, and interjection. Training and development sets containing audio snippets and their corresponding labels were provided to participants at the beginning of the challenge.

For Track I, the submitted systems are evaluated based on stuttering event detection accuracy, recall, precision, and F1 score on the audio snippets in the test set. The F1 score, which is the harmonic mean of precision and recall, is calculated as follows:
\begin{align}
F1 = 2 \cdot \frac{\text{Precision} \cdot \text{Recall}}{\text{Precision}+\text{Recall}} .
\end{align}

\subsubsection{Track II-Automatic Speech Recognition}
The main goal of Track II is to advance the development of ASR systems that can effectively handle stuttering speech. Participants must devise speech-to-text systems that accurately recognize speech containing stuttering events, converting it into clean text with the stuttering event labels removed. Training and development sets containing stuttering speech audio and their corresponding text transcriptions were provided for system development.

For Track II, the accuracy of the ASR system is measured by Character Error Rate (CER). The CER indicates the percentage of characters that are incorrectly predicted. It calculates the minimum number of insertions (Ins), substitutions (Subs), and deletions (Del) required to transform the hypothesis output into the reference transcript. Specifically, CER is calculated as follows:
\begin{align}
    CER=\frac{N_{\text {Ins }}+N_{\text {Subs }}+N_{\text {Del }}}{N_{\text {Total }}} \times 100 \% ,
\end{align}
where$N_{\text{Ins}}$, $N_{\text{Subs}}$ and $N_{\text{Del}}$ represent the number of insertion, substitution, and deletion errors, respectively, and $N_{\text{Total}}$ is the total number of characters in the reference transcript. As standard practice, insertions, deletions, and substitutions all contribute to the overall error rate.

\subsubsection{Track III Research Paper Track}
Participants are invited to contribute research papers that utilize the stuttering speech dataset and evaluation framework in their experimental setups and analyses. This track provides an opportunity to explore and document innovative approaches and findings related to stuttering speech technologies.

% There are no specific evaluation metrics for Track III. We welcome paper submissions that use stuttering speech data to advance research in this field.

\section{System Description}
\subsection{Baseline system}
We provide a Conformer~\cite{conformer} baseline for SED track and a U2++~\cite{21u2++} baseline for ASR track, built using the WeNet toolkit~\cite{wenet}. These baselines serve as benchmarks to facilitate reproducible research and comparison across different submissions.

\begin{figure}[ht]
\centering
\includegraphics[width=1.0\linewidth]{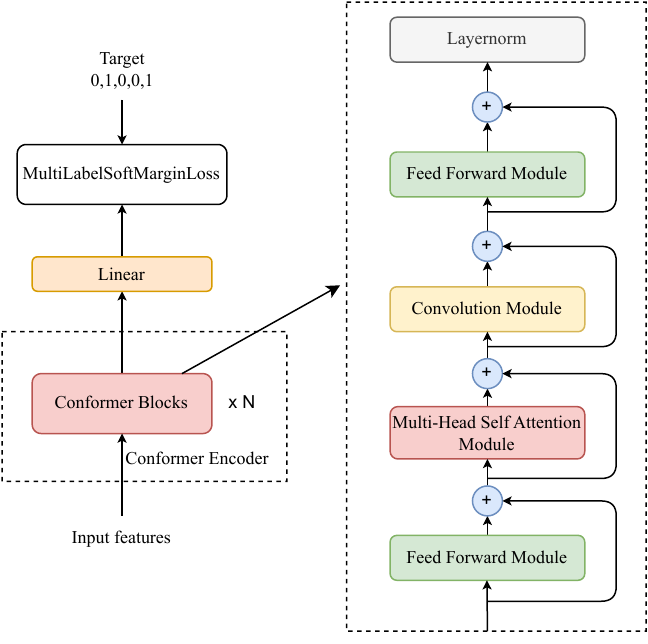}
\caption{\centering Baseline model structure in Track I. Target 0,1,0,0,1 indicates the presence or absence of the five stuttering events.}
\label{fig:sed_model}
\vspace{-10pt}
\end{figure}

\subsubsection{Track I-Stuttering Event Detection}
The baseline system consists of multiple Conformer blocks, as depicted in Fig~\ref{fig:sed_model}. Each block comprises a multi-head self-attention module, a convolution module, and two feed-forward modules, with 4 attention heads and an output size of 256. The convolutional modules use a kernel size of 15 to capture a wide range of temporal features. Due to the limited training data, we use 3 Conformer blocks, resulting in a parameter size of 9.7 million.

We first extract 80-dimensional filter bank (fbank) features from the speech, feed them into the Conformer encoder, and perform classification through a linear layer. The model employs single-task learning to predict the five stuttering event types, using the multi-label soft margin loss for training. The model is optimized using the MultiLabelSoftMarginLoss function in PyTorch.
% the ground truth:
% \begin{align}
%     L=MultiLabelSoftrMarginLoss(y, y^-),
% \end{align}

We train the model using only the AS-70 dataset. The training is conducted with an initial learning rate of 0.001 on four NVIDIA 4090 GPUs, with a batch size of 16 per GPU. The model is trained for 100 epochs, with 1,000 warmup steps.

\subsubsection{Track II-Automatic Speech Recognition}
\begin{figure}[ht]
\centering
\includegraphics[width=1.0\linewidth]{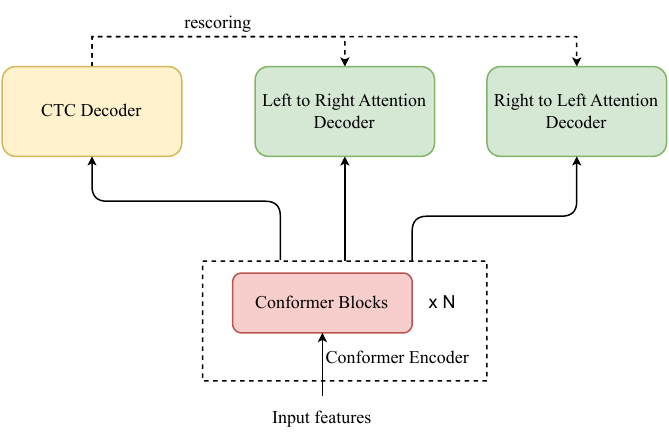}
\caption{\centering Baseline U2++ model structure in Track II.}
\label{fig:asr_model}
\vspace{-10pt}
\end{figure}

The U2++ model~\cite{21u2++}, illustrated in Fig~\ref{fig:asr_model}, is a unified two-pass framework with bidirectional attention decoders. It incorporates future contextual information through a right-to-left attention decoder to enhance the representational ability of the shared encoder and improve performance during the rescoring stage. The baseline model for Track II consists of 12 Conformer encoder layers, each with 4 attention heads and an output size of 256. The convolutional modules have a kernel size of 8. Additionally, the baseline model includes 3 bitransformer decoder layers, with 3 left-to-right and 3 right-to-left layers.

We use both the AS-70 dataset and the AISHELL-1 dataset for training. The model is trained with an initial learning rate of 0.001 on four NVIDIA 4090 GPUs, with a batch size of 16 per GPU. It is trained for 100 epochs and 25,000 warmup steps.

\subsection{Submission of Track I}
Track I receives a total of 24 modeling results from 5 teams. We have selected the top 3 teams for presentation, in addition to the official system model.

\noindent\textbf{Official system: }
We aim to enhance the performance of the Conformer model~\cite{conformer} in SED through various simple data augmentation techniques. Specifically, we incorporate speed perturbation~\cite{ko2015audio} and data balancing to improve the robustness and generalization of the model. Speed perturbation alters the playback speed of audio samples without changing the pitch. A fixed set of perturbation factors 0.8-1.2 are introduced in the speech samples, where the intervals are 0.05. This technique introduces variability into the training data, enabling the model to recognize stuttering events under different temporal conditions.
Additionally, we observe significant differences in the number of utterances for each stuttering event as shown in Table~\ref{tab:sed_data}. To address data imbalance, we employ oversampling techniques to balance the dataset. By ensuring equal representation of stuttering and non-stuttering events, we aim to mitigate the model's bias towards the majority class, thereby improving detection accuracy.

\noindent\textbf{T018: }
Team T018 improves the performance of the SED task through several data augmentation techniques and the introduction of the Zipformer-L architecture~\cite{24zipformer}. 
They implement a comprehensive data augmentation strategy to simulate Mandarin stuttered speech using the Montreal Forced Aligner (MFA)~\cite{mcauliffe2017montreal} trained on the AISHELL-1 dataset~\cite{aishell_2017}. Six types of stuttering data are generated: block, sound repetition, word repetition, block with sound repetition, block with word repetition, and normal speech. The quantities for each type are specified to balance the dataset. The augmentation process involved alignment, segmentation, augmentation, combination, and validation steps.
The model is based on the Zipformer-L architecture~\cite{24zipformer}, a variant of the Transformer~\cite{17attention} designed explicitly for speech processing tasks. The model consists of several key components for the SED task: an input layer of filter bank features, Zipformer layers, and a classification layer for the five stuttering events.

\noindent\textbf{T029: }
Team T029 introduces a novel SED model which combines the Conformer model~\cite{conformer} with Bidirectional Long Short-Term Memory (BiLSTM) networks~\cite{20bilstm}. The Conformer model extracts acoustic information from stuttered speech, while the BiLSTM networks capture contextual relationships within the speech. This combination simplifies the acquisition of both speech signal representations and contextual relations for stuttering patients. Additionally, T029 explores the impact of multitask settings on the model's performance to address the issue of imbalanced data. They design a model comprising five binary classification tasks for each stuttering type, with the output layer defined as 5*2 dimensions.

\noindent\textbf{T031: }
Team T031 proposes a Fine-Grained Contrastive Learning (FGCL) framework for SED. This method aims to improve the accuracy of stutter detection by capturing subtle nuances that previous SED techniques have overlooked. FGCL leverages a detailed acoustic analysis to model frame-level probabilities of stuttering events. The team then introduce a novel mining algorithm to identify both easy and confusing frames within audio data. By applying a stutter contrast loss~\cite{20contrastive}, they refine the representation of these frames, thereby enhancing the model's ability to distinguish between stuttered and fluent speech.

% Likelihood Modeling: The likelihood of detecting stuttering instances in each frame is modeled using a classifier to obtain frame-level Class Activation Scores (CAS), which are then used to derive the frame-level probabilities of stuttering events.

% Confusing Frames Mining: A mining algorithm is introduced to identify confusing frames within the audio. This algorithm samples confusing frames through cascaded contraction and expansion operations on pseudo-boundaries, while easy frames are identified using top-k and bottom-k selection methods.

% Stutter Contrast Loss: A stutter contrast loss (LSC) is proposed to refine the representation of confusing frames, thereby enhancing discrimination between stuttered and non-stuttered speech. This loss function includes two components: Lst for refining confusing stuttered features and Lnst for refining confusing non-stuttered features.

\subsection{Submission of Track II}
Track II receives 73 modeling results submitted by 6 teams, and we have selected the top 3 teams for presentation.

\noindent\textbf{T006: } 
Team T006 focus on enhancing stuttering ASR through comprehensive data augmentation techniques. Their innovative approach includes both signal-based and adversarial methods to extend the variety and volume of speech data available. \textbf{Speed Perturbation}~\cite{ko2015audio}: Adjust the time-domain speech signal's sampling resolution, creating variations in speech rate and volume to replicate the speech rate variations typical of PWS. \textbf{Tempo Perturbation}: Utilize the WSOLA algorithm~\cite{93wosla} to alter speech duration while maintaining pitch, simulating the slower speech cadence observed in PWS. \textbf{InsertSilence}: Randomly insert silent segments into the speech signal, effectively introducing pauses and delays that mimic stuttering blocks, aiding the ASR model in recognizing and processing speech patterns with frequent pauses. \textbf{RepeatPart}: Repeat parts of the audio signal multiple times, simulating the repetitive articulations often seen in stuttering. \textbf{SpecAugment}~\cite{park2019specaugment}: Apply dynamic masks to spectral features, enhancing the model's ability to generalize by simulating partial information loss and small speech fragments. \textbf{GAN-Based Augmentation}: Utilize Generative Adversarial Networks like Parallel WaveGAN, BigVGAN, and Vocos~\cite{20wavegan, 20vocos} to inject fine-grained spectral-temporal variations, capturing detailed differences between stuttering and non-stuttering speech. These multi-stage data enhancement techniques generated 1,443 hours of ASR data. For the model architecture, the team use a 17-layer E-Branchformer encoder~\cite{22E-branchformer} and a 6-layer transformer decoder, implemented using the ESPnet toolkit~\cite{18Espnet}.

% T006 employs a multi-stage data augmentation strategy aimed at accurately recognizing stuttered speech which includes both signal-based and adversarial methods to extend the variety and volume of speech data available for training. These methods include speed perturbation, tempo perturbation, insert silence, repeat part, specAugment, and GAN-based data augmentation. 

% The core methods involve: Simulating the overall changes in rhythm, speech rate, and spectral envelope typical in PWS by inserting silent segments, repeating parts of the speech, and modifying the speech rate and rhythm of healthy speech samples.
% Using GAN-based data augmentation~\cite{20wavegan, 20vocos} to introduce fine-grained spectral temporal features associated with stuttering speech, thereby enabling the ASR system to better learn and recognize the nuances of stuttered speech.

\begin{table}[t!]
\centering
\caption{SED results (F1 scores) in Track I.}
\label{tab:sed_res}
\begin{tabular}{@{}lcccccc@{}}
\toprule
TeamID   & \multicolumn{1}{l}{/p} & \multicolumn{1}{l}{/b} & \multicolumn{1}{l}{/r} & \multicolumn{1}{l}{\textbf{[]}} & \multicolumn{1}{l}{/i} & \multicolumn{1}{l}{Avg} \\ \midrule
Baseline & 65.12                  & 24.3                   & 41.86                  & 61.85                   & 74.87                  & 53.6                    \\ 
Offical     & 64.67                  & 36.70                  & 57.61                  & 65.81                   & 79.38                  & 60.83                    \\
T018     & 67.61                  & 31.58                  & 59.25                  & \textbf{81.11}                   & 83.94                  & 64.7                    \\
T029     & \textbf{70.89}                  & \textbf{44.07}                  & \textbf{59.78}                  & 74.79                   & \textbf{85.15}                  & \textbf{66.93}                   \\
T031     & 65.89                  & 24.78                  & 50.15                  & 64.85                   & 77.06                  & 56.55                   \\
\bottomrule
\end{tabular}
\end{table}

\noindent\textbf{T018: }
Team T018 enhance the performance of the ASR task using the same data augmentation techniques applied in the SED task, combined with a different model architecture. The ASR model is based on the RNN-T architecture~\cite{12rnnt}, with the speech encoder utilizing the Zipformer-L~\cite{24zipformer}. This model architecture comprises the following components: an input layer for filter bank features, a Zipformer encoder, a transducer decoder, and an output layer.

\noindent\textbf{T051: }
Team T051 leverage the ESPnet toolkit~\cite{18Espnet} to integrate a CTC and attention-based encoder-decoder network. The encoder utilizes the Branchformer architecture~\cite{22branchformer}. To enhance the model's generalization ability and robustness, the team employs SpecAugment~\cite{park2019specaugment} for data augmentation during training.

% in tests, significantly outperforming the baseline model's 19.18% CER. The team's efforts earned them third place in the SLT 2024 competition.
% The encoder adopts the Branchformer architecture~\cite{22branchformer}, designed for efficient processing of speech signals. It features relative self-attention mechanisms and employs a conditional gated multi-layer perceptron, aiming to capture both local and global context in speech.
% To enhance the model's generalization ability and robustness, SpecAugment is used for data augmentation during training. This involves manipulations like time warping, frequency masking, and time masking, enriching the diversity of training data without the need for additional recorded samples.

\begin{table*}[t!]
\centering
\caption{ASR results (CER\%) in Track II.}
\label{tab:asr_res}
\begin{tabular}{@{}lcccccc@{}}
\toprule
TeamID   & \textbf{Mild} & \textbf{Moderate} & \textbf{Server} & \textbf{Conversation} & \textbf{Command} & \textbf{Avg} \\ \midrule
Baseline & 17.27                   & 19.45                       & 29.53                     & 17.7                            & 21.5                      & 19.18     \\ 
% T005     & 16.08                   & 18.12                       & 27.96                     & 16.92                           & 19.49                     & 17.92     \\
T006     & \textbf{11.16}                   & \textbf{12.38}                       & \textbf{18.54}                     & \textbf{12.13}                           & \textbf{12.57}                     & \textbf{12.30}      \\
T018     & 15.36                   & 16.07                       & 25.8                      & 14.29                           & 20.84                     & 16.85     \\
T051     & 15.63                   & 18.01                       & 26.13                     & 16.44                           & 18.7                      & 17.33     \\
\bottomrule
\end{tabular}
\end{table*}

\section{CHALLENGE RESULTS SUMMARY}
This section presents the results of the StutteringSpeech Challenge, focusing on Track I and Track II. We provide a detailed analysis of the leaderboard outcomes, highlighting the top-performing teams and their methodologies.

\subsection{Results and Analysis of Track I}
\noindent\textbf{Overall: }
The leaderboard results for Track I are shown in Table~\ref{tab:sed_res}. Across both results, It is evident that data augmentation techniques play a critical role in improving performance across various models. 
Notably, the Conformer-BILSTM model of T029, equipped with five classification heads, achieves the highest mean F1 scores.
The Zipformer model of T018 demonstrates effective detection in word or phrase repetition events due to its robust content information capturing capability.
The FGCL model of T031 outperforms the Conformer model in terms of average F1 score even without data augmentation.

\noindent\textbf{T029: }
Team T029 achieves the highest mean F1 scores across the five stuttering events, with notable performance in four individual events. Their success is attributed to their innovative Conformer-BILSTM model, effectively capturing contextual information after extracting local stuttering acoustic features. Additionally, they use five classification heads to detect each stuttering event separately, simplifying the detection process and enhancing accuracy.

\noindent\textbf{T018: } 
Team T018 secures the second-highest average F1 score, with an improvement of 20.7\% compared to the baseline. They achieves the highest score in the \textbf{[]} event, significantly outperforming other models in this category. The Zipformer model demonstrates its efficacy in speech recognition tasks due to its robust content information capturing proficiency. \textbf{[]} is word or phrase repetition, the Zipformer exhibits effective detection in this aspect.

\noindent\textbf{T031: } 
Team T031 finishes third overall. Their FGCL model showed a 5.5\% improvement in the average F1 score compared to the baseline model. It is worth highlighting that T031 did not use data augmentation but focuses solely on model modifications, contributing significantly to the development of SED. Combining their model with data augmentation techniques could enhance performance further.

\noindent\textbf{Official: } 
Our official system demonstrates a 13.5\% improvement in the average F1 score compared to the baseline. While not participating in the rankings, our model, which incorporates only speed perturbations and data balancing, showed notable improvements. Speed perturbation, in particular, is highly effective, likely because stuttering is closely related to speech speed, and altering the speed generates diverse stuttering data.

\noindent\textbf{Stuttering Types: }
Various data augmentation methods can enhance the accuracy of SED. \textbf{/p} involves sound prolongation, while \textbf{/r} is characterized by sound repetition. \textbf{[]} represents word repetitions. These events are closely related to the duration of speech, which makes speed perturbation a particularly method for enhancing the diversity and accuracy of SED in the official system. \textbf{/b} involves blocking, and \textbf{/i} involves interjections, which can be augmented by inserting silent segments or unnatural interjections, thereby generating more diverse data, as exemplified in T018. 

\subsection{Results and Analysis of Track II}

\noindent\textbf{Overall: }
The leaderboard results are presented in Table~\ref{tab:asr_res}. 
It is evident that data augmentation techniques also play a critical role in improving ASR performance.
Team T006's comprehensive augmentation strategies are particularly noteworthy, especially their enhancements for voice commands. Future tasks on stuttering voice commands can benefit from T006's approach.
Additionally, the E-branchformer and Branchformer models demonstrates superior performance compared to the Conformer model in stuttering ASR tasks.

\noindent\textbf{T006: }
Team T006 achieves first place in all test sets. Their average CER is 27\% lower than the second-place team and 35.87\% lower than the baseline model. This can be attributed to their comprehensive data augmentation techniques, including signal-based and adversarial-based augmentations. Their use of the E-branchformer architecture also outperforms Conformer model in ASR task.

\noindent\textbf{T018: } 
Team T018 secures second place in Track II, with an average CER reduction of 12.1\% compared to the baseline. Their success is partly due to their data augmentation techniques and utilizing the powerful Zipformer model structure for the ASR task. Although their data augmentation is slightly less effective compared to T006, the potential applicability of the Zipformer structure in stuttering ASR tasks warrants further exploration.

\noindent\textbf{T051: }
Team T051 finishes third, with an average CER reduction of 9.6\% compared to the baseline. This indicates that the Branchformer structure is more effective in stuttering ASR tasks.

\noindent\textbf{Stuttering Severity: } 
In all outcomes, CER improvements with the severity of stuttering. None of the participating teams specifically addresses stuttering severity in their models. Future research should focus on developing ASR systems that can better handle varying levels of stuttering severity.

\noindent\textbf{Conversation and Command: } 
The results show a notable difference in CER between command and conversation scenarios, with the exception of T006. T006 significantly reduces this gap, suggesting that their data augmentation methods improves the recognition rate for command speech, making it more promising for voice wake-up applications in PWS.

% \subsection{Analysis}
% Across both tracks, it is evident that data augmentation techniques played a critical role in improving performance. Team T029's novel Conformer-BILSTM model and Team T006's comprehensive augmentation strategies are particularly noteworthy. Future research should explore the integration of these successful techniques to further enhance ASR and SED systems for stuttering speech.

% Additionally, the impact of stuttering severity on CER highlights the need for models specifically designed to handle varying levels of stuttering severity. The differences in CER between command and conversation scenarios also suggest potential areas for improvement, particularly in voice wake-up applications for individuals who stutter.

\section{conclusion}
The StutteringSpeech Challenge has significantly advanced both ASR and SED systems tailored for PWS. By leveraging the comprehensive AS-70 dataset, the challenge has demonstrated the effectiveness of various data augmentation techniques and novel model architectures. Notably, in the SED task, the Conformer-BILSTM model achieved the highest mean F1 scores, while the Zipformer model excelled in detecting word repetitions.
For the ASR task, comprehensive augmentation strategies were particularly noteworthy, especially their enhancements for voice commands. The E-branchformer and Branchformer models demonstrated superior performance in stuttering ASR tasks.
These findings underscore the importance of tailored approaches in both stuttering SED and ASR tasks and highlight the potential of different data augmentation techniques. Future research should focus on integrating these models with real-world applications, enhancing their robustness to different stuttering severities. By continuing to refine these technologies, we can develop more reliable and accessible speech recognition solutions, ultimately improving the quality of life for PWS.

% \clearpage
\bibliographystyle{IEEEbib}
\bibliography{refs}

\end{document}